\documentclass[12pt,english]{article}
\usepackage{titling}
\thanksmarkseries{alph}
\usepackage[round]{natbib}
\usepackage{inputenc,epsfig, amsmath, amsthm,graphicx, amssymb, graphics,color}
\title{Stylized facts of the Indian Stock Market}
\author{Rituparna Sen and Manavathi S}
\date{}

\begin{document}
\maketitle
\begin{abstract}
Historical daily data for eleven years of the fifty constituent stocks of the NIFTY index traded on the National Stock Exchange have been analyzed to check for the stylized facts in the Indian market. It is observed that while some stylized facts of other markets are also observed in Indian market, there are significant deviations in three main aspects, namely leverage, asymmetry and autocorrelation. Leverage and asymmetry are both reversed making this a more promising market to invest in. While significant autocorrelation observed in the returns points towards market inefficiency, the increased predictive power is better for investors.
\end{abstract}
{\it{JEL classification}}: C58,G15,C55\\
{\it{AMS classification}}: 62P05, 91B80, 91B84,62M10. \\
{\bf{Keywords:}} National Stock Exchange, Volatility Clustering, Leverage Effect, Heavy Tails, Power Law.

\pagebreak
\section{Introduction}
Any stock market deals with the trading of shares of stocks which involves inherent uncertainty. Because of the huge amounts of money involved, economists use different tools and measures to develop models for stock prices. These models are then used for portfolio selection, risk management, derivative pricing etc. It becomes important to analyze and test the accuracy of the model before putting it to use. There are some properties of stocks which are common across markets, irrespective of the nature of the stock. They have been identified after observing that a large volume of data, across markets and time periods, follow these properties. These properties are known as the stylised facts.  They provide a starting point for economic models of the stock price. It is reasonable to expect the prices arising from any acceptable model to follow at least these properties.

Stylized facts are well studied in the literature. The paper \citet{cont01}, on the empirical properties of asset returns lists such stylized facts. The current literature on the presence and identification of these facts  is reviewed in \citet{thompson13}.  The stylized facts that are observed frequently are heavy tails, volatility clustering, slow decay of auto-correlation in absolute returns and leverage effect. Absence of simple arbitrage, power law decay of the tails of the return distribution and volatility clustering are studied in \citet{cristelli14}. Three models used for modelling and forecasting volatility, namely, the standard GARCH, Exponential GARCH, and the Autoregressive Stochastic Volatility model are studied in \citet{malmsten10}. It is found that none of the models dominate the others when it comes to reproducing stylized facts.

There has been comparatively less work in stylized facts in the emerging markets and in particular on the Indian market. As it is seen in other aspects of market behaviour, the emerging markets often behave very differently from the developed markets. For example, in contrast to the stylized fact about heavy tails, the KOSPI index of the Korean stock market was found to follow an exponential distribution, see \citet{oh06}. The trend of investments is accelerating in the Indian market as a result of regulatory reforms and removal of other barriers for the international equity investments. This increased momentum is expected to show up in the price dynamics. In India, the emergence and growth of derivative market is relatively a recent phenomenon. Since its inception in June 2000, derivatives market has exhibited exponential growth both in terms of volume and number of traded contracts. Also, NSE is pure order driven market. Most markets like NYSE and NASDAQ are a hybrid of order and quote driven. For all these reasons, it is important to explore, on a large scale, the behavior of the Indian market. \citet{kumar18} studies the liquidity aspect of the Indian market and finds a lot of commonality. \citet{mukherjee11} studies the stylized facts based on only the BSE SENSEX index. Although the index is a good proxy for the market, it is not sufficient to capture the nature of the individual stocks.

In this paper, daily data for eleven years on fifty individual stocks traded in the National Stock Exchange (NSE) have been analyzed. Eight stylized facts out of the listed eleven in \citet{cont01} are studied because the remaining three involve intra-day data. Statistical package R is used for the analysis.

The rest of the paper is organized as follows. In section 2 the data set is described. Section 3 explores simple distributional properties of log returns, namely asymmetry, normality and leverage. Section 4 studies the time series properties of returns, squared returns and absolute returns. In sections 5 and 6, heavy tails and conditional heavy tails of the series are studied. In each section we state the stylized fact and present the inferences, along with technical details on how the analysis is done. Section 7 concludes the work that is done and offers possible areas to apply these findings.

\section{Description of data}
\begin{figure}[h]\begin{center}	
		\includegraphics{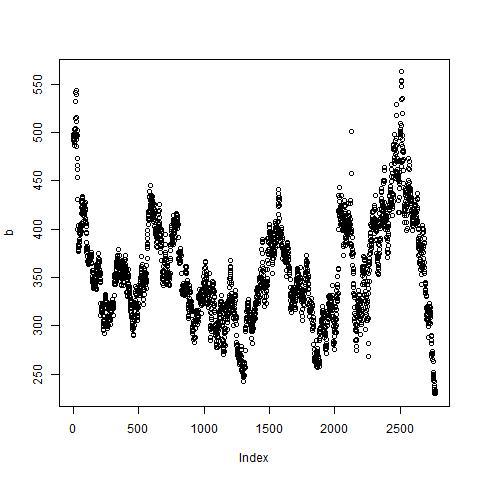}
	\end{center}
	\caption{Prices of Bosch}\label{fig:Bosch}
\end{figure}

\begin{figure}[h]\begin{center}	
		\includegraphics{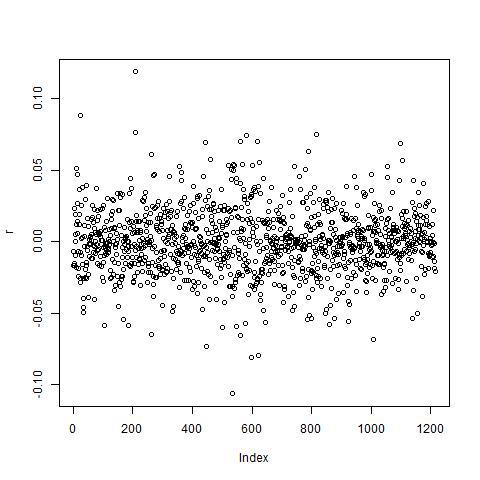}
	\end{center}
	\caption{Log Returns of BPCL}\label{fig:BPCL}
\end{figure}

The historical data is downloaded from investing.com.  Fifty stocks that are traded in the National Stock Exchange(NSE) are considered. The stocks are the constituents of the NIFTY index. For each stock the data contains seven attributes for each trading day, namely, the date, closing price, opening price, the highest price, lowest price, volume traded and the change percentage on that day. These were taken over the time range of January 2007 to November 2017. The plot of prices of the Bosch Ltd(BOSH) stock is shown in Figure \ref{fig:Bosch}.
Since the prices are often non-stationary, it is more common to use log returns for statistical analysis. The log return of a stock $R(t)$ at time $t$ is given by \[R(t) = \log(S(t))-\log(S(t-1)) \] where \( S(t) \) denotes price of stock at time $t$. For the rest of the paper we deal with the log return and not the prices. For illustrative purpose, the log returns for the Bharat Petroleum Corporation Limited (BPCL) stock are shown in Figure \ref{fig:BPCL}. A visual comparison with Figure \ref{fig:Bosch} shows that the concern of non-stationarity has been addressed.

\section{Simple Distributional properties}
In this section we explore the simple stylized facts related to the distribution of returns, ignoring time series dependence and tail behaviour. In particular, we study the symmetry and normality of the distribution, as well as the inter-dependence of the return and volatility.
\subsection{Gain loss asymmetry}
A common stylized fact is the gain loss asymmetry.  One observes large drawdowns in stock prices and stock index values but not equally large upward movements.
The skewness of a random variable $X$ is the third standardized moment and is denoted by $\gamma_1$. \[ \gamma_1 = \frac{E(X-\mu)^3}{(\sqrt{E(X - \mu)^2})^\frac{3}{2}}, \]
where $\mu=E(X)$ is the mean of the distribution. Skewness measures the asymmetry of the probability distribution of a random variable. A positive skew distribution means that the right tail is longer than the left tail. A negative skew distribution means that the left tail is longer. Large drawdowns compared to upward movements would correspond to a long left tail and hence, a negatively skewed curve.

\begin{figure}[h]\begin{center}\hspace{-1in}
		\includegraphics[width=.2\textwidth,trim=0.5in 4.5in 2.5in 0.2in]{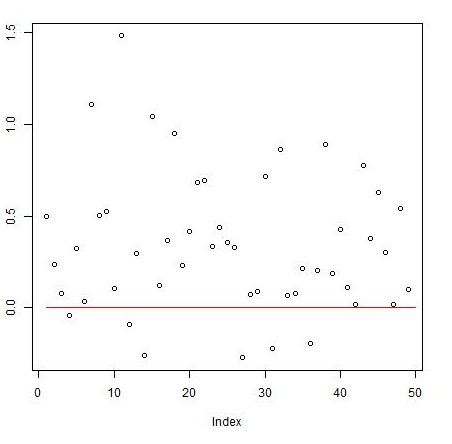}
	\end{center}
	\vspace{2.5in}
	\caption{A plot of the skewness values.}\label{fig:Skewness}
\end{figure}

For the data set under consideration, skewness of the returns was calculated. Most stocks have positive skewness as shown in Figure \ref{fig:Skewness}.

It is observed that most stocks have positive skewness and thus show larger upward movements than drawdowns. This is in contrast with the stylized facts reported in \citet{cont01}. In \citet{huang14}, the authors compare asymmetry indices of historical prices from ten stock markets using market index data. They find that in most stock markets, price fall is faster than price rise; while in China and India, price rise is generally faster than price fall. Our results reconfirm this for the Indian market at the stock level.
\subsection{Aggregational Gaussianity}
The next stylized fact is aggregated Gaussianity, which is the following phenomenon. As one increases the time scale $\Delta$t over which returns are calculated, their distribution looks more and more like a normal distribution. In particular, the shape of the distribution is not the same at different time scales.

\begin{figure}[h]\begin{center}\hspace{-3in}
		\includegraphics[width=.4\textwidth,trim=0in 3.5in 2.5in 0.5in]{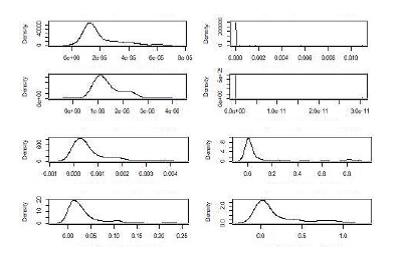}
	\end{center}
	\vspace{4in}
	\caption{Kernel density plots of p-values of SW (left) and KS (right) tests for daily, weekly, monthly and quarterly (from top to bottom) returns.}\label{fig:Normality}
\end{figure}

Two normality tests in R are performed on the data, namely the Kolmogorov-Smirnov (KS) test and the Shapiro-Wilke (SW) test for daily, weekly, monthly and quarterly returns. The p-values increase as the time over which the returns are calculated increases. The kernel density plots of these p-values for the 50 stocks under consideration are presented in Figure \ref{fig:Normality}.

It is seen that as the time scale increases, the p-vales are less and less concentrated around zero and the normality assumption is rejected for fewer stocks. Thus, the distribution of returns of many stocks become similar to the normal distribution as the time over which returns are calculated is lengthened. We conclude that aggregated Gaussianity is present in the data.

\subsection{Leverage Effect}
Leverage effect refers to the observation that most measures of volatility of an asset are negatively correlated with the returns of that asset. If there is high volatility in the stock movement, then the returns will be low. The rationale behind this is that if there is a lot of fluctuating movement of the stock price, not many investors will invest in the stock.

\begin{figure}[h]\begin{center}
		\includegraphics[width=.15\textwidth,trim=2.5in 3.5in 2.5in .5in]{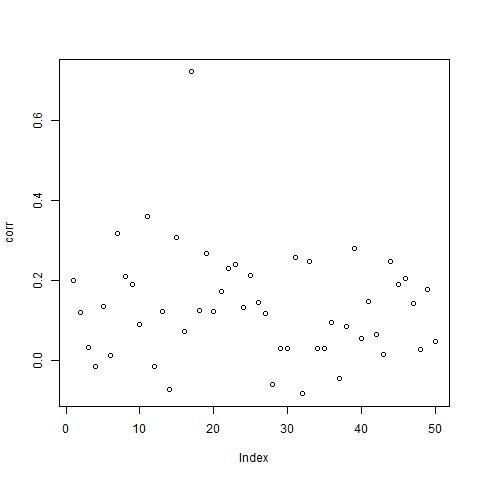}
	\end{center}
	\vspace{1.5in}
	\caption{Correlation between returns and volatility.}\label{fig:Leverage}
\end{figure}

The correlation between returns and squared returns is calculated.  Figure \ref{fig:Leverage} shows a density plot (for all fifty stocks) of the correlation between volatility and returns. It is expected to be negative, but is found to be positive. Thus the observed results are contrary to the fact stated in \citet{cont01}.

\section{Time Series Dependence}
In this section we explore the time series properties of the returns. In particular, we look at the autocorrelation of returns, absolute returns and squared returns. We start with some definitions.
The autocorrelation of a stationary time series $X_t, t=1, \cdots, T$, denoted by $\rho_X$ is defined as a function of the lag as. \[ \rho_X (k) = cor(X_s, X_{s+k}) = \frac{E{(X_s - \mu_X)(X_{s+k} - \mu_X)}}{\sigma^2_X}. \] Here $\mu_X$ and $\sigma^2_X$ are respectively the mean and variance of $X_s$. Due to stationarity, $\mu_X$, $\sigma^2_X$ and $\rho_X (k)$ do not depend on $s$.
The autocorrelation of a time series measures the linear correlation between the original time series ($X_s$) and the lagged series ($X_{s+k}$). The autocorrelation depends on the lag $k$. For example, the autocorrelation of a time series X with the daily returns of a stock with lag 1 would indicate how much a day's return influences the next day's return, or how much a day's return depends on the previous day's return.

The auto-covariance $\gamma_X(k)$ of a time series $X$ is given by \[ \gamma_X (k) = cov{\{X_s , X_{s+k}\}} = E{{(X_s - \mu_X)((X_{s+k} - \mu_X))}} \]

Partial autocorrelation $\phi_X(k)$ measures the influence of the value at a given instant on the value at the instant at lag $k$, controlling for the effect of intermediate values. That is,\[\phi_X(k)=cor(X_s,X_{s+k}|X_{s+1}, \cdots, X_{s+k-1}).\]

\subsection{Autocorrelation of returns}
The stylized fact is that autocorrelations of asset returns are often insignificant, except for very small intraday time scales for which microstructure effects come into play. This fact is the reason why investing in stocks is risky. This is why it is difficult to predict the future stock prices. If the returns were correlated, then it would mean that the return values are dependent on previous return values, and the correlation coefficient can be used to determine the expected value of the future return and hence price.

\begin{figure}[h]\begin{center}	
		\includegraphics[width=.65\textwidth]{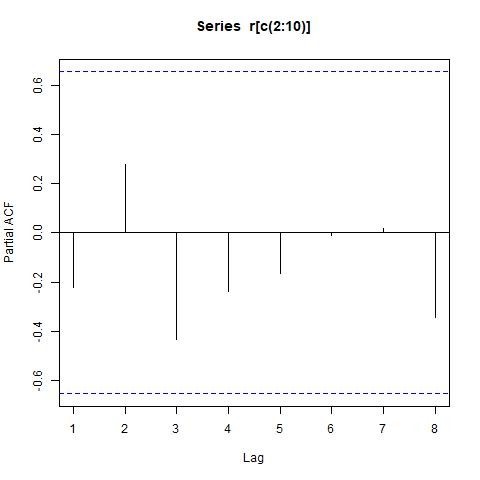}
	\end{center}
	\caption{Partial autocorrelation values for MRTI.}\label{fig:MRTI}
\end{figure}

Here the autocorrelation and partial autocorrelation for all the 50 series has been calculated for the returns of the stocks using a lag of 10. These quantities are estimated using the method of moments estimators. For demonstrative purpose, Figure \ref{fig:MRTI} shows the partial autocorrelation (PACF) values for the Maruti Suzuki India Ltd (MRTI) stock. The dashed lines mark the 5\% significance levels. Observe that all partial autocorrelation upto order 10 are insignificant.

For a formal evaluation, the Portmanteau tests are carried out to test the hypothesis that all autocorrelations upto a certain lag are zero. These tests are to determine lack of fit of the data. That is, they determine how close the data are to white noise. It gives a measure of how much one value depends on the previous value(s). There are two Portmanteau tests: Box-Pierce and Ljung-Box. Both tests give similar results. So we present the results for the Ljung-Box test only.

The Ljung-Box test statistic is given by
\begin{equation}Q =n(n+2) \sum_{k=1}^m \frac{\hat\rho_k^2}{n-k}\end{equation}
Here this variable $Q$ is follows a $\chi^2$  distribution with $m$ degrees of freedom under the null hypothesis. Here $n$ is the total number of observations,  $m$ is the maximum lag up to which autocorrelation is determined.

Using a maximum lag of 10 and level of significance 1\%, the null hypothesis is rejected for 22 stocks out of 50. This implies that there is some autocorrelation in many of the stocks under consideration. This is again at variance with the observations from other markets and can potentially be useful to predict future prices, giving rise to arbitrage opportunities.

\subsection{Volatility Clustering}\label{sec:Vol}
Different  measures  of  volatility  display  a  positive  autocorrelation  over  several  days, which quantifies the stylized fact that high-volatility events tend to cluster in time. If there is  volatility clustering,  it means that there is  some significant autocorrelation in the volatility of the stocks, so that the values depend on the previous values and tend to be similar over time.  So if the volatility on one day is high, then that will cause the volatility to remain high over the subsequent days. One measure of the volatility of a stock is the variance of the returns.  On calculating the average return ($\mu$) for all the stocks, most stocks had average return close to zero.  So for this data set the squared returns can be considered as a fairly accurate measure of volatility.

\begin{figure}[h]\begin{center}	
		\includegraphics[width=.65\textwidth]{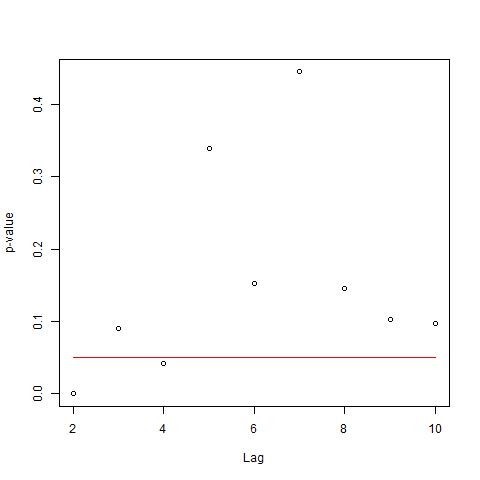}
	\end{center}
	\caption{p-values for ACF of squared returns of HDFC.}\label{fig:HDFC}
\end{figure}

The autocorrelation of squared returns was calculated for a lag upto 10 and formal hypothesis tests are conducted using absence of autocorrelations between squared returns as null hypothesis and test statistic: $X=\frac{\sqrt{n} \hat{\rho}}{1-\hat{\rho}^2}$ where $n$ is the number of observations, and $\hat{\rho}$ is the estimated autocorrelation. Under the null hypothesis, $X$ follows a standard normal distribution asymptotically. This can be used to compute the p-value of the test. Majority of p-values were less than 5\%. The plots of p-values of the squared returns of HDFC Bank (HDFC) stock is shown in Figure \ref{fig:HDFC}.
Portmanteau tests were also conducted. All the stocks show a non-zero autocorrelation between the squared returns. The autocorrelation at lag one is positive in almost all cases.

In summary, the data does display volatility clustering, as expected.

\subsection{Slow Decay of Autocorrelation in Absolute Returns}
The autocorrelation function of absolute returns decays slowly as a function of the time lag, roughly as a power law with an exponent $\beta\in [0.2, 0.4]$. This stylized fact means that the effect of absolute returns on the future values does not ’wear off’ soon. This is sometimes interpreted as an indication of long-range dependence.

To find the tail index, a power law has to be fitted for autocorrelation with the lag values. A linear regression is fitted for the logarithm values of autocorrelation ($ac$) and lag ($l$). The power law will be of the form $ac = k l^\alpha$, for some proportionality constant $k$. Taking logarithm on both sides,
\begin{equation}\log(ac) =  \alpha \log(l) + \log(k)\end{equation}
The slope of the regression line is the exponent of the power law.  The negative reciprocal of this exponent gives the tail index.

\begin{figure}[h]\begin{center}	
		\includegraphics[width=.65\textwidth]{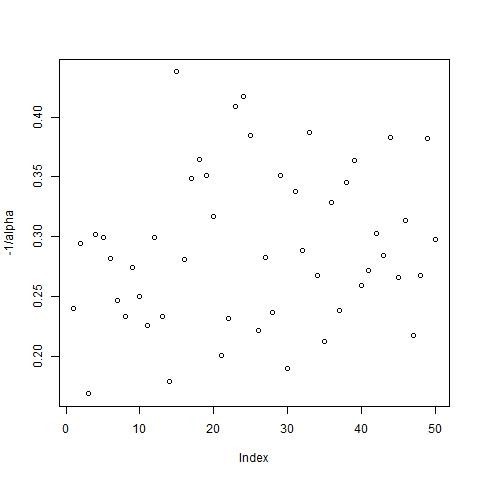}
	\end{center}
	\caption{Plot of tail index.}\label{fig:Decay}
\end{figure}

The tail index of autocorrelation with lag was computed. The values are shown in Figure \ref{fig:Decay}. The stocks do have slow decay of autocorrelation with the exponent in the expected range.

\section{Heavy Tails}\label{sec:HT}
The stylized fact regarding tails states that the distribution of returns seems to display a power-law or Pareto-like tail, with a tail index which is finite, higher than two and less than five.

A power law is a functional relationship between two variables, where the relative change in one variable is proportional to the relative change in the other. It is of the form \[ Y = kX^\alpha \] where $X$ and $Y$ are variables of interest, $\alpha$ is the law's exponent and $k$ is a constant. Pareto Law is a special power law, also known as distributional power law, where $Y$ is the probability involving a random variable. For example, \[ P(S>x) = \frac{k}{x^\zeta} \] is a Pareto Law. The exponent $\zeta$ is independent of the units in which the law is expressed.

Distributions of random variables are studied in comparison with the exponential distribution. The tail of a distribution is the part of the distribution where $|X|$ tends to $\infty$. The thickness of the tail is the tail index. Distributions can be classified as being heavy tailed or light tailed. A heavy-tailed distribution has a tail that is not bounded by the exponential tail, whereas the light-tailed distribution has a tail that falls below the exponential tail. Here we look only at the tail and not the part of the distribution before where the tail begins. The choice of the point in the distribution where the tail begins is also important in determining tail index. Consider any distribution $P(X)$ with cumulative distribution function $F(x)=1−\overline {F}(x)$ defined by $Pr (X>x)= \overline {F}(x)$, such that for some $\xi$ >0,
\[\overline {F}(x)=x^\frac{-1}{\xi} L(x)\]
where $L(x)$ is some slowly varying function for large $x$.
The tail index of the fat-tailed distribution $P(X)$ is by definition $\xi$.

\begin{figure}[ht]\begin{center}	
		\includegraphics[width=.6\textwidth]{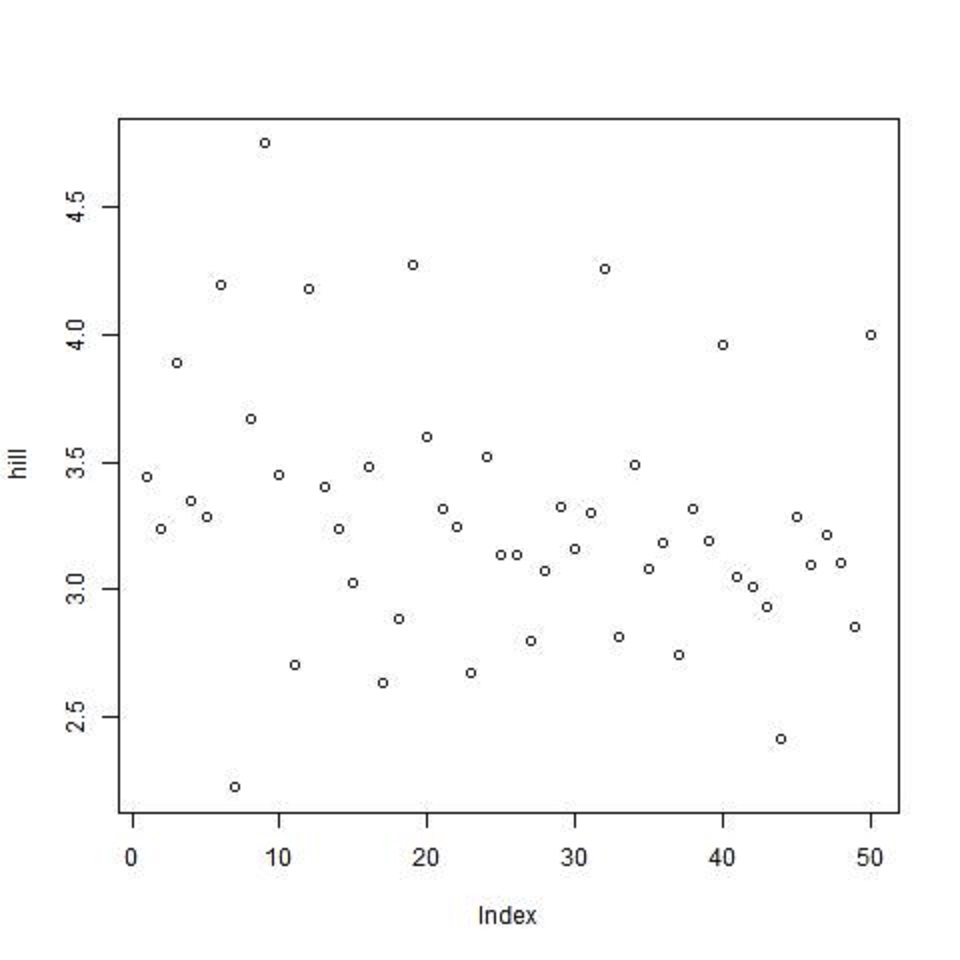}
	\end{center}
	\caption{Tail Index of returns.}\label{fig:Hill}
\end{figure}

Using the hill.adapt() function in the extremefit package of R, the tail index is calculated. The values for all 50 stocks are shown in Figure \ref{fig:Hill}. It is observed that the returns indeed have heavy tails and the index is found to lie between two and five.

\section{Conditional Heavy Tails}
The final stylized fact that we look at is conditional heavy tails. Even after correcting returns for volatility clustering (e.g.  via GARCH-type models), the residual time series still exhibit heavy tails. However, the tails are less heavy than in the unconditional distribution of returns.

As seen in section \ref{sec:Vol} there is volatility clustering in financial time series. This can lead to heavy tails of the return distribution observed in section \ref{sec:HT}. One approach to avoid the heavy tails is to model the volatility process itself as a time series to capture the clustering behaviour of volatility. This is generally achieved though ARCH and GARCH models.

The general time series model in statistical analysis is:
\begin{equation}X_t=\mu_t+\epsilon_t.\end{equation} Here, $\mu_t$ is the conditional mean of the series, that is, $\mu_t = \mathrm{E}[X_t|X_{t-1}, X_{t-2}, \cdots]$ and $\epsilon_t$ is a disturbance term. In traditional time series analysis, the disturbance term is usually assumed to be a White Noise innovation, and the conditional mean is expressed as a function of past observations:
\begin{equation} X_t = \alpha_0 + \alpha_1X_{t-1} + \cdots + \alpha_pX_{t-p} +\epsilon_t, \end{equation}
where $\epsilon_t$ is a White Noise innovation. Under these assumptions, the conditional mean of $X_t$ is non-constant and time dependent, but the conditional variance is a fixed quantity and equal to the marginal variance. In other words, there is some short-term memory in the mean, but not in the variance. We are seeking an enhanced formulation that allows for non-constant conditional variance. One possibility is to decompose the disturbance term as $\epsilon_t=\sigma_t W_t$, where $\sigma_t$ is the conditional standard deviation and $W_t$ is a White Noise innovation. If it is assumed that $\sigma_t$ is a function of previous instances, we obtain a process that also has short-term memory in the variance and hence can be used for volatility modeling. Because $\sigma_t$ is a (conditional) standard deviation, it needs to be made sure that the function of previous instances is non-negative. That is cumbersome to achieve with linear combinations, because coefficient restrictions are always awkward. Instead, it is more popular to work with non-linear variance function models. One such model is the ARCH (Autoregressive Conditional Heteroscedastic) model. A series $\epsilon_t$ is said to follow a first-order autoregressive conditional heteroscedastic process,or short, is ARCH(1), if \begin{equation}\epsilon_t = \sigma_t W_t \quad\mathrm{with}\quad \sigma_t^2 = \beta_0 + \beta_1\epsilon^2_{t-1}.\end{equation} Here $W_t$ is a White Noise innovation process, with mean zero and unit variance. The two parameters are the model coefficients. There are some disadvantages of the ARCH model. The model assumes that positive and negative shocks have the same effects on volatility. In practice, it is well known that asset prices respond differently to positive and negative shocks. It is restrictive in the sense that $\beta_1$ should be in($0, 1/\sqrt{3}$) for a finite fourth moment. Thus a large number of extensions of the standard ARCH model have been suggested, see \citet{engle12} for a review. For a GARCH($p,q$) model (generalized autoregressive conditional heteroscedasticity) the variance is given by
\begin{equation} \sigma^2_t=\beta_0 + \beta_1\epsilon^2_{t-1}+\cdots+\beta_q\epsilon^2_{t-q}+\gamma_1\sigma^2_{t-1}+\cdots+\gamma_p\sigma^2_{t-p}\end{equation}
The following constraints are required on the parameters.
For positivity we need $\beta_0 > 0; \beta_1, \cdots,\beta_{q-1}\ge 0; \beta_q > 0; \gamma_1,\cdots,\gamma_{p-1}\ge 0;\gamma_p>0$
and for stationarity we need
$\beta_1 + \cdots + \beta_q + \gamma_1 + \cdots+ \gamma_p < 1$.

\begin{figure}[h]\begin{center}
		\includegraphics[width=.15\textwidth,trim=2.5in 3.5in 2.5in .5in]{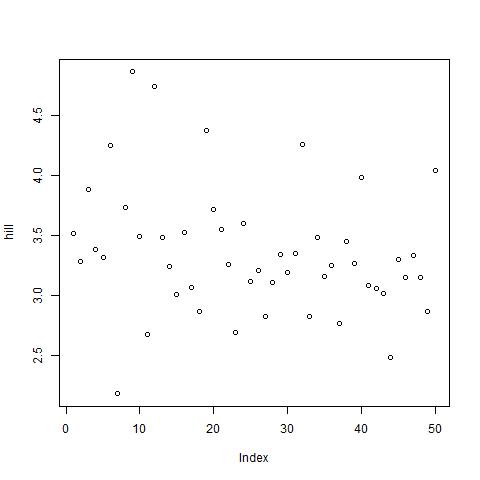}
	\end{center}
	\vspace{1.5in}
	\caption{Tail Index of GARCH fitted returns.}\label{fig:TailGARCH}
\end{figure}

Tail  index  is  found  for  GARCH-fitted  residuals. The  tails  are  still  heavy  and GARCH fitting seems to have no effect on the estimated tail index.  This is shown in Figure \ref{fig:TailGARCH}.

\section{Conclusion}
Stylized facts for a large number of stocks of the Indian stock market were studied. After a brief survey of the literature on stylized facts, the data is described. Then,  the  facts  were analyzed using the basic functions and toolboxes provided in R. Based on the observations made, certain inferences are drawn and stated.

Among the accepted stylized facts, the ones that are found in the Indian market are aggregated Gaussianity, volatility clustering, power law decay of autocorrelation of absolute returns with exponent in the rage 0.2 to 0.4, heavy tails with an index between 2 and 5. In particular this excludes stable laws with infinite variance and the normal distribution. However the precise form of the tails is difficult to determine.

There are some significant deviations of the Indian data from the stylized facts listed for developed markets. We have found difference in three aspects: leverage, autocorrelation and asymmetry. Also, the tail-index is not reduced by GARCH-fitting in most cases. The significance of these differences are discussed separately below.

The gain loss asymmetry in reversed from what is seen in developed markets. For the Indian market the distribution of returns is positively skewed most stocks show larger upward movements than drawdowns. The same phenomenon is also observed for European emerging markets in \citet{karpio07}. This is advantageous for investors as the waiting time for a specific amount of gain is shorter than that for the same amount of loss.

The leverage effect is reversed for most stocks, that is, the returns and volatility are positively correlated. The leverage effect was first discussed by \citet{black76} who observed that the volatility of stocks tends to increase when the price drops. Empirical evidence of negative leverage has been documented widely, for eg., see \citet{bouchaud01}. Asymmetric GARCH models have been developed to capture negative leverage and feedback effects. In the Indian market, we see positive leverage. Hence, there is a need for new explanations and new models. Since both the leverage effect and gain/loss asymmetry share many common features, \citet{ahlgren07} attempts to link them together.

22 out of 50 stocks show significant autocorrelation in the returns. Autocorrelation of returns points towards market inefficiency. More detailed study needs to be done in the lines of  \citet{poterba88}. It is necessary to find the cause and direction of auto-correlation. For eg, large proportion of noise trading and short-term investors may lead to mean-reversion. For investors, autocorrelation in return is better than random walk as past returns have predictive power over future returns and can be captured and used for investment decisions.

This analysis can be used to decide what to test for in a predictor model for the Indian data.  Several new investment techniques can be devised using the leverage, asymmetry and autocorrelation, that are different from those used in other markets.
\setcitestyle{numbers}
\bibliographystyle{agsm}
\bibliography{refs}
\end{document}